\documentclass[apj]{emulateapj}
\usepackage{graphics}

\slugcomment{Accepted version}

\shorttitle{Black hole powered nebulae and the case of IC342 X-1}
\shortauthors{Cseh et al.}

\begin{document}

\title{Black Hole Powered Nebulae and a Case Study of the Ultraluminous X-ray Source IC342 X-1}

\author{D\'{a}vid Cseh\altaffilmark{1}, St\'{e}phane Corbel\altaffilmark{1}, Philip Kaaret\altaffilmark{2}, Cornelia Lang\altaffilmark{2}, Fabien Gris\'{e}\altaffilmark{2},\\ 
Zsolt Paragi\altaffilmark{3,4}, Anastasios Tzioumis\altaffilmark{5}, Valeriu Tudose\altaffilmark{6} and Hua Feng\altaffilmark{7}}
\email{david.cseh@cea.fr}

\altaffiltext{1}{Laboratoire Astrophysique des Interactions Multi-echelles (UMR 7158), CEA/DSM-CNRS-Universite Paris Diderot, CEA Saclay, F-91191 Gif sur Yvette, France}
\altaffiltext{2}{Department of Physics and Astronomy, University of Iowa, Van Allen Hall, Iowa City, IA 52242, USA}
\altaffiltext{3}{Joint Institute for VLBI in Europe, Postbus 2, 7990 AA Dwingeloo, The Netherlands}
\altaffiltext{4}{MTA Research Group for Physical Geodesy and Geodynamics, PO Box 91, 1521 Budapest, Hungary}
\altaffiltext{5}{Australia Telescope National Facility, CSIRO, PO Box 76, Epping, NSW 1710, Australia}
\altaffiltext{6}{ASTRON, Oude Hoogeveensedijk 4, 7991 PD Dwingeloo, The Netherlands}
\altaffiltext{7}{Department of Engineering Physics and Center for Astrophysics, Tsinghua University, Beijing 100084, China}

\begin{abstract}

We present new radio, optical, and X-ray observations of three Ultraluminous X-ray sources (ULXs) that are associated with large-scale nebulae. We report the discovery of a radio nebula associated with the ULX IC342 X-1 using the Very Large Array (VLA). Complementary VLA observations of the nebula around Holmberg~II X-1, and high-frequency Australia Telescope Compact Array (ATCA) and Very Large Telescope (VLT) spectroscopic observations of NGC5408 X-1 are also presented. We study the morphology, ionization processes, and the energetics of the optical/radio nebulae of IC342 X-1, Holmberg II X-1 and NGC5408 X-1. The energetics of the optical nebula of IC342 X-1 is discussed in the framework of standard bubble theory. The total energy content of the optical nebula is $6\times10^{52}$ erg. The minimum energy needed to supply the associated radio nebula is $9.2\times10^{50}$~erg. In addition, we detected an unresolved radio source at the location of IC342 X-1 at VLA scales. However, our Very Long Baseline Interferometry (VLBI) observations using the European VLBI Network likely rule out the presence of any compact radio source at milli-arcsecond (mas) scales. Using a simultaneous Swift X-ray Telescope measurement, we estimate an upper limit on the mass of the black hole in IC342 X-1 using the "fundamental plane" of accreting black holes and obtain $M_{\rm{BH}}\leq(1.0\pm0.3) \times10^{3}$~M$_{\odot}$. Arguing that the nebula of IC342 X-1 is possibly inflated by a jet, we estimate accretion rates and efficiencies for the jet of IC342 X-1 and compare with sources like S26, SS433, IC10 X-1.
 
\end{abstract}


\section{Introduction}

Ultraluminous X-ray sources are variable non-nuclear X-ray sources in external galaxies with luminosities greatly exceeding the Eddington luminosity of a stellar-mass compact object, assuming isotropic emission \citep{CM,phil01}. The irregular variability, observed on time scales from seconds to years, suggests that ULXs are binary systems containing a compact object that is either a stellar-mass black hole with beamed \citep{king,elmar} or super-Eddington emission \citep{beg}; or an intermediate-mass black hole (IMBH). IMBHs have been invoked in contexts ranging from the remnants of Population III \citep{mad} stars to the formation of supermassive black holes (SMBHs) \citep{ptak}; SMBHs may form through the hierarchical merger of lower mass black holes \citep[eg.][]{ebi}.

Several ULXs show emission-line optical nebulae, which can be used as a calorimeter to infer the total intrinsic power of the ULX \citep{pakull3,pakull2,dave}. In general, the nebulae around ULXs are either photoionized due to the high X-ray and UV luminosity of the compact object \citep{pakull3,pakull2,hoii, philneb,moon}, or shock-ionized driven by jets, outflows, and/or disk winds \citep{pakull3,rob,abol}. In several cases, two-component optical line profiles are present, indicating a mixture of the two mechanisms or alternatively the narrow line could be due to the shock precursor. Another common feature is the presence of the high-ionization He {\sc ii} emission line. It can have various origins: the nebula, the donor, the accretion disk, or a disk wind. The ionizing photon rate needed to produce the narrow high-excitation He {\sc ii} line of the nebulae in photoionized sources indicates that their X-ray emission is at most mildly beamed \citep{pakull3,pakull2,hoii,philneb,moon}. 

Only a handful of radio detections of ULXs have been made so far, including NGC5408 X-1 \citep{phil,soria,cornelia}, Ho II X-1 \citep{miller}, and MF16 \citep{dyk,laci}, if not considered a supernova remnant \citep{mato}. Most of these sources show large nebulae ($>\sim$50~pc) that are likely powered by continuous energy input from the ULX. Shock-dominated ones are probably powered in the same manner as the W50 nebula is powered by the Galactic binary SS433 \citep{dubner}. However, the ULX radio nebulae require greater total energy content than W50. A similarly powerful nebula, S26, was found by \citet{pakull} in optical and  \citet{s26} in radio. For other possible radio associations with ULXs, we refer the reader to \citet{sanchez, sor,mez}.

The X-ray spectra of ULXs share some similar properties with the canonical black hole states of Galactic black hole binaries (GBHBs). A number of ULXs show state transitions \citep{kubota,feng0,feng,olivier,fab0,math}. In GBHBs during the X-ray hard state, the sources are associated with self-absorbed compact jets \citep{steph,robfender}. Given the similarities between ULXs and GBHBs, it is interesting to investigate the presence of such compact jets for ULXs with hard X-ray spectra.

In this paper, we present new radio and optical observations of two ULXs that are associated with large-scale nebulae: Holmberg II X-1 (Ho II X-1) and NGC5408. In addition, we present discovery of a radio nebula associated with IC342 X-1. In Section 2, we describe observations, data analysis and results. The energetics of the optical and radio nebulae and the jet properties of IC342 X-1 are investigated in Section 3.  In Section 4, we briefly summarize our results.

\section{Observations, Analysis, and Results}

\subsection{VLA observations of IC342 X-1 and Ho II X-1}

Observations of IC342 X-1 and Ho II X-1 were carried out using the C- and B-array configurations of the Very Large Array (VLA) of the National Radio Astronomy Observatory (NRAO). The observations were made at 4.8 and 8.5 GHz (VLA program code: AL711) and the details are summarized in Table 1. Data calibration, combination in the (u,v) plane and imaging were carried out using the NRAO Astronomical Image Processing System (AIPS: e.g., \citet{greis}). We adjusted the Robust weighting parameter between 0 and -2 to bring out the fine-scale radio emission and parameters are identified in image captions. 

\begin{table*}
\begin{center}
\caption{Summary of Observations\label{obs}}
\begin{tabular}{lcccccc}
\tableline\tableline
Source & Instrument & Config. &Central& On-source&Observation&Flux\\
&&&Frequency&Time&Date&\\
\tableline
IC342 X-1 & VLA & B & 4.8 GHz & 3.2 h& 6 Dec 07&2.0$\pm0.1$ mJy\\
IC342 X-1 & VLA & C & 4.8 GHz & 3.2 h& 25 Apr 08&*\\
Ho II X-1 & VLA & B & 4.8 GHz & 3.2 h& 8 Dec 07&613$\pm61 \mu$Jy\\
Ho II X-1& VLA & C& 8.5 GHz& 3.2 h& 21 Apr 08&395$\pm 40 \mu$Jy\\
\tableline
NGC5408 X-1 & ATCA & 6D & 5.5 GHz & 12 h & 23 Aug 09&226$\pm33\mu$Jy\\
NGC5408 X-1 & ATCA & 6D & 9 GHz & 12 h& 23 Aug 09&137$\pm 36 \mu$Jy\\
NGC5408 X-1 & ATCA & 6D & 17.9 GHz & 12 h& 23 Aug 09&76$\pm 20 \mu$Jy\\
\tableline
NGC5408 X-1 & ESO VLT & - & 575-731 nm & 0.7 h& 12 Apr 10&-\\
\tableline
IC342 X-1  & EVN & - &  1.6 GHz & 12 h & 15 Jun 11&$\leq21 \mu$Jy**\\
 IC342 X-1 &  Swift/XRT &  PC&  0.3 - 10 keV & 9.4 ks& 15 Jun 11&2.67***\\
\end{tabular}
\tablecomments{*B and C configuration data were combined.**Three-sigma upper limit.***Unabsorbed flux in units of 10$^{-12}$ erg cm$^{-2}$ s$^{-1}$.}
\end{center}
\end{table*}

\subsubsection{Radio detection of IC342 X-1}

\begin{figure*}
\begin{center}
\rotatebox{0}{
\includegraphics[width=6in]{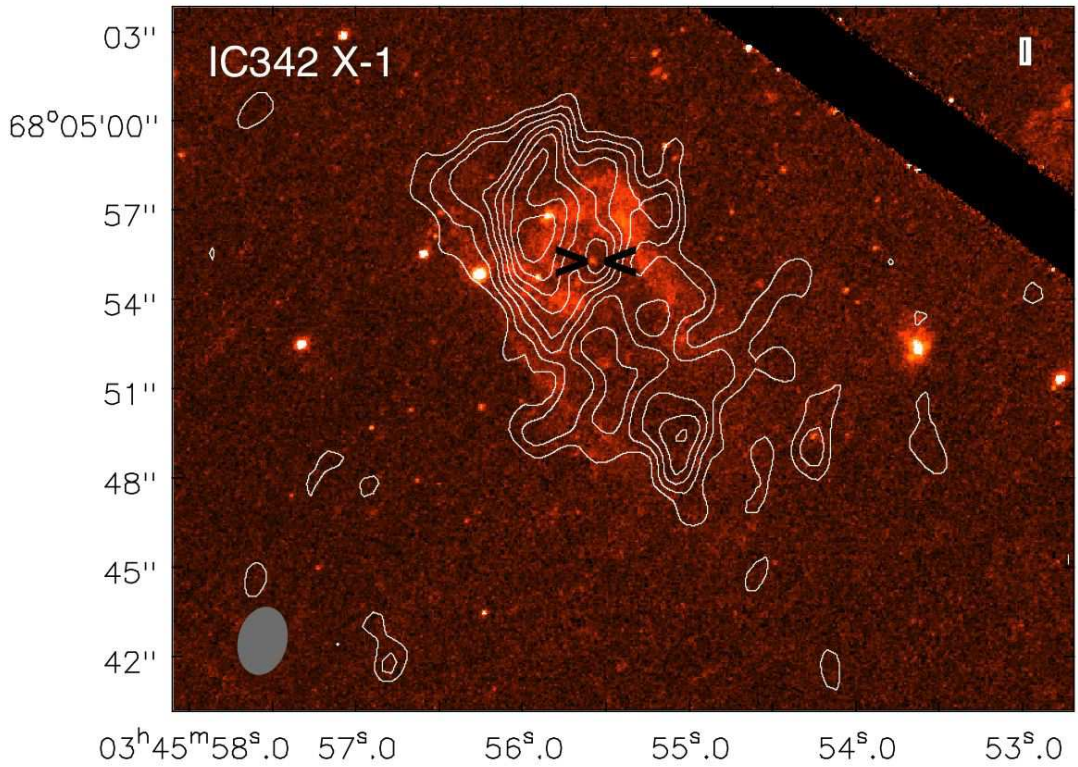}}
\end{center}
\caption{VLA 5 GHz image of IC342 X-1 overlaid on the H$\alpha$ HST image. Contours represent radio emission and are drawn at 3, 4, 5, 6, 7, 8, 9, and 10 times the rms noise level of 11 $\mu$Jy beam$^{-1}$. The peak brightness is 122.4 $\mu$Jy beam$^{-1}$. The resolution of the image is $2\farcs3 \times 1\farcs6$ at PA=$-13\degr$ and the image was made with Robust=0 weighting. The sign '$>\,<$' marks the X-ray position of the ULX.}
 \label{ICFOV}
 \end{figure*}

Figure~\ref{ICFOV} shows the 5 GHz VLA B- and C-array combined image of IC342 X-1 with Robust=0 weighting overlaid on the H$\alpha$ HST image of \citet{fengIC}. Extended radio emission is present surrounding the position of IC342 X-1.  Diffuse emission is detected at up to the 10-$\sigma$ level with peak intensity of $\sim$120 $\mu$Jy beam$^{-1}$ and the estimated total flux density in the nebula is $\sim2$~mJy. The corresponding luminosity at a distance of 3.9~Mpc \citep{dist} is $L_{neb}=1.8 \times 10^{35}$ erg s$^{-1}$; with $L= \nu L_{\nu}$. We find that the size of the nebula is 16" $\times$ 8", which corresponds to 300 pc $\times$ 150 pc. We estimate the volume of the nebula by taking a sphere with a diameter of 12". Comparing the 5 GHz radio map to the H$\alpha$  image, the optical and the radio nebulae are both elongated in the NE--SW direction, possibly exhibiting a shock-front. The size of the radio nebula is consistent with the size of the optical nebula \citep{pakull3,fab00,fengIC}.

\begin{table*}
\begin{center}
\caption{Hubble images and source parameters\label{Hobs}}
\begin{tabular}{lcccccc}
\tableline\tableline
Name & Narrow Band Filter & Centered on & Ref.&Distance&Optical&Radio\\
&&&&&diameter&diameter\\
\tableline
IC342 X-1 & F658N & H$\alpha$ & 1&3.9 Mpc&190 pc&225 pc\\
Ho II X-1 & FR462N & He {\sc ii} & 2&3.39 Mpc&45 pc&60 pc\\
Ho II X-1 & FR505N & H$\beta$ & 2&-&101 pc&-\\
NGC5408 X-1 & F502N & [O {\sc iii}]$\lambda$5007 & 3&4.8 Mpc&60pc&40 pc\\
\end{tabular}
\tablecomments{References: 1 - \citet{fengIC}, 2 - \citet{hoii}, 3 - \citet{fab}} 
\end{center}
\end{table*}

\begin{figure*}
\begin{center}
\rotatebox{0}{
\includegraphics[width=7in]{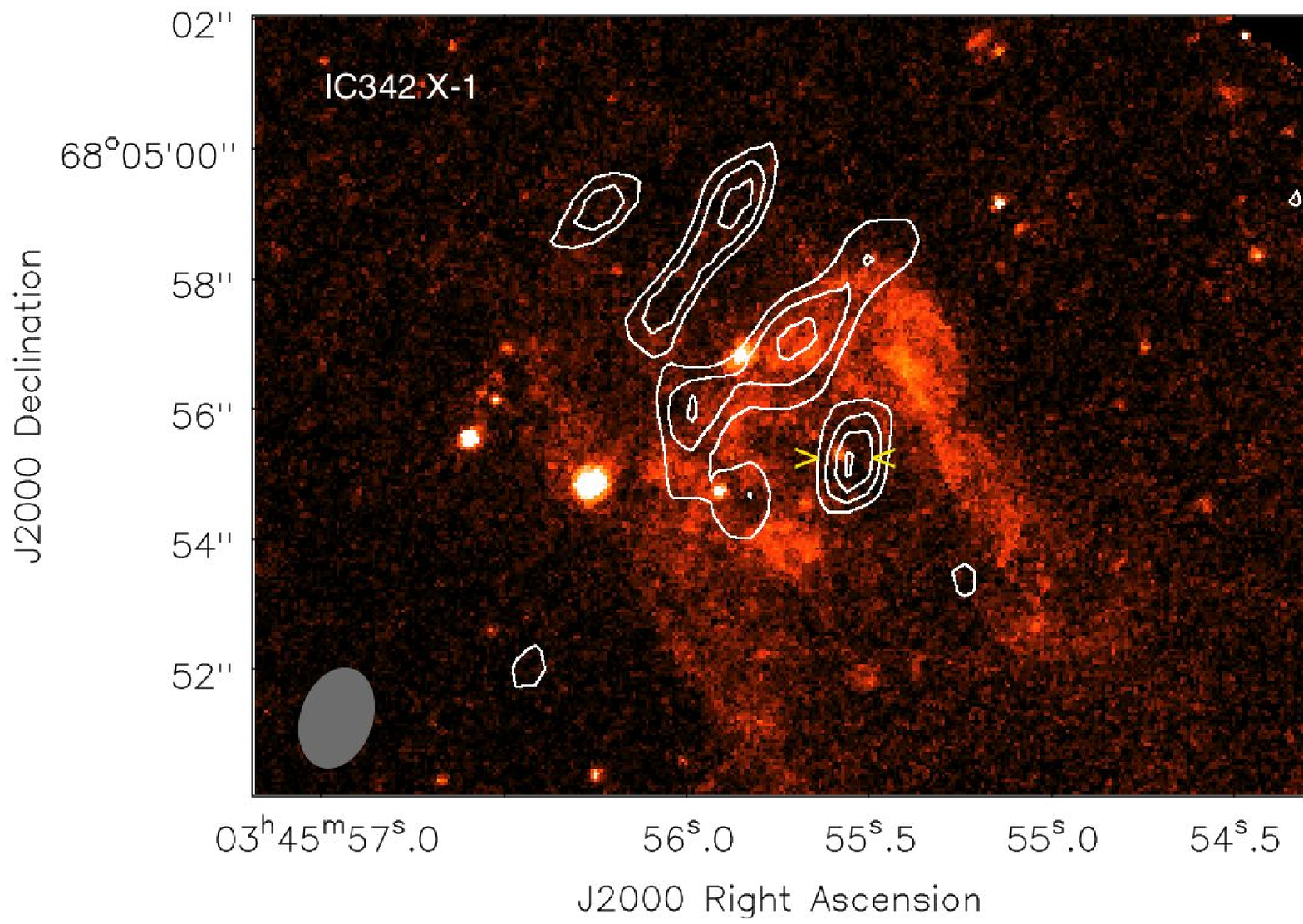}}
\end{center}
\caption{VLA 5 GHz image of IC342 X-1 overlaid on the H$\alpha$ HST image shown in Figure~\ref{ICFOV}. This image was made with Robust=-2 weighting and therefore the largest extended features are missing. The resolution is $1\farcs6 \times 1\farcs1$ at PA=$-19\degr$. Contours represent radio emission and are drawn at 3, 4, 5, and 6 times the rms noise level of 15 $\mu$Jy beam$^{-1}$. The peak brightness is 96.3 $\mu$Jy beam$^{-1}$. The sign '$>\,<$' marks the X-ray position of the ULX.}
 \label{IC}
 \end{figure*}

Figure~\ref{IC} illustrates a more uniformly weighted image of the 5 GHz radio emission surrounding IC342 X-1. We weighted the radio image with a robust parameter of -2 in order to resolve out the diffuse nebular emission and show the fine scale structure. The strongest radio emission appears towards the NE of the ULX and is coincident with the strongest H$\alpha$ emission.  The radio emission extends farther to the NE in regions of little or no H$\alpha$ emission.

Figure~\ref{IC} also reveals an unresolved radio source at the location of IC342 X-1. This unresolved source is detected with a flux density of $\sim96.3 \, \mu$Jy at the 6.5-$\sigma$ level. The corresponding luminosity is $8.8 \times 10^{33}$ erg s$^{-1}$. The HST position of the ULX is RA=03h45m55.61s, Decl.= +68\degr04'55.3" (J2000.0) with the 90\% positional errors of 0.2 arcsec \citep{fengIC}. We obtained the position of the radio peak of RA=03h45m55.54s, Decl.= +68\degr04'55.18" using the {\sc maxfit} task of AIPS. The distance between the two positions is $\sim$0.4 arcsec. Using the AIPS task  {\sc jmfit}, we estimate a radio positional uncertainty of $\sim$0.1 arcsec at one sigma. Therefore the optical and radio positions are consistent using 90\% positional errors. 

\subsubsection{Multi-frequency radio observations of Ho II X-1}

\begin{figure*}
\rotatebox{0}{
\includegraphics[width=3.5in]{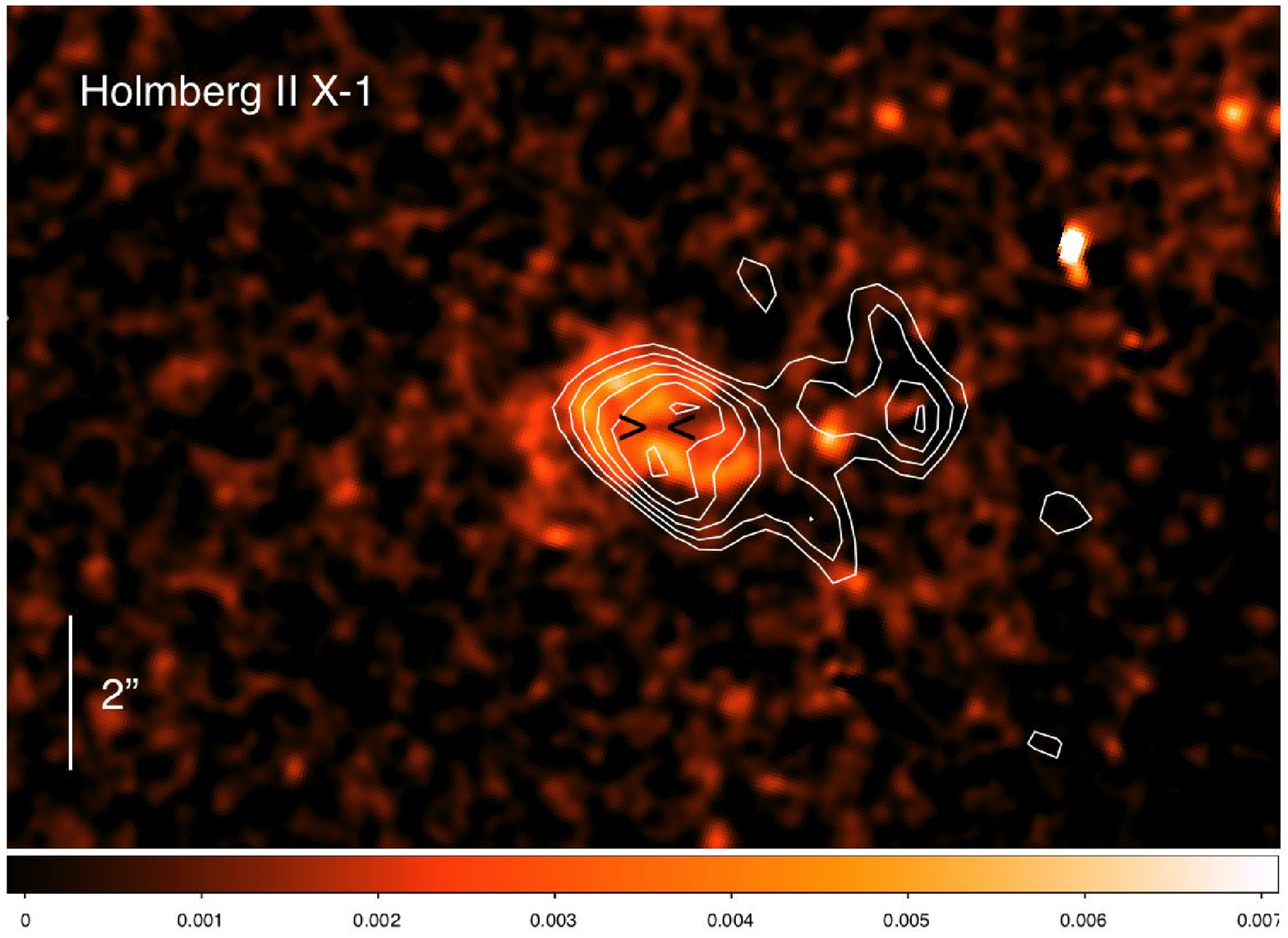}}
\rotatebox{0}{
\includegraphics[width=3.5in]{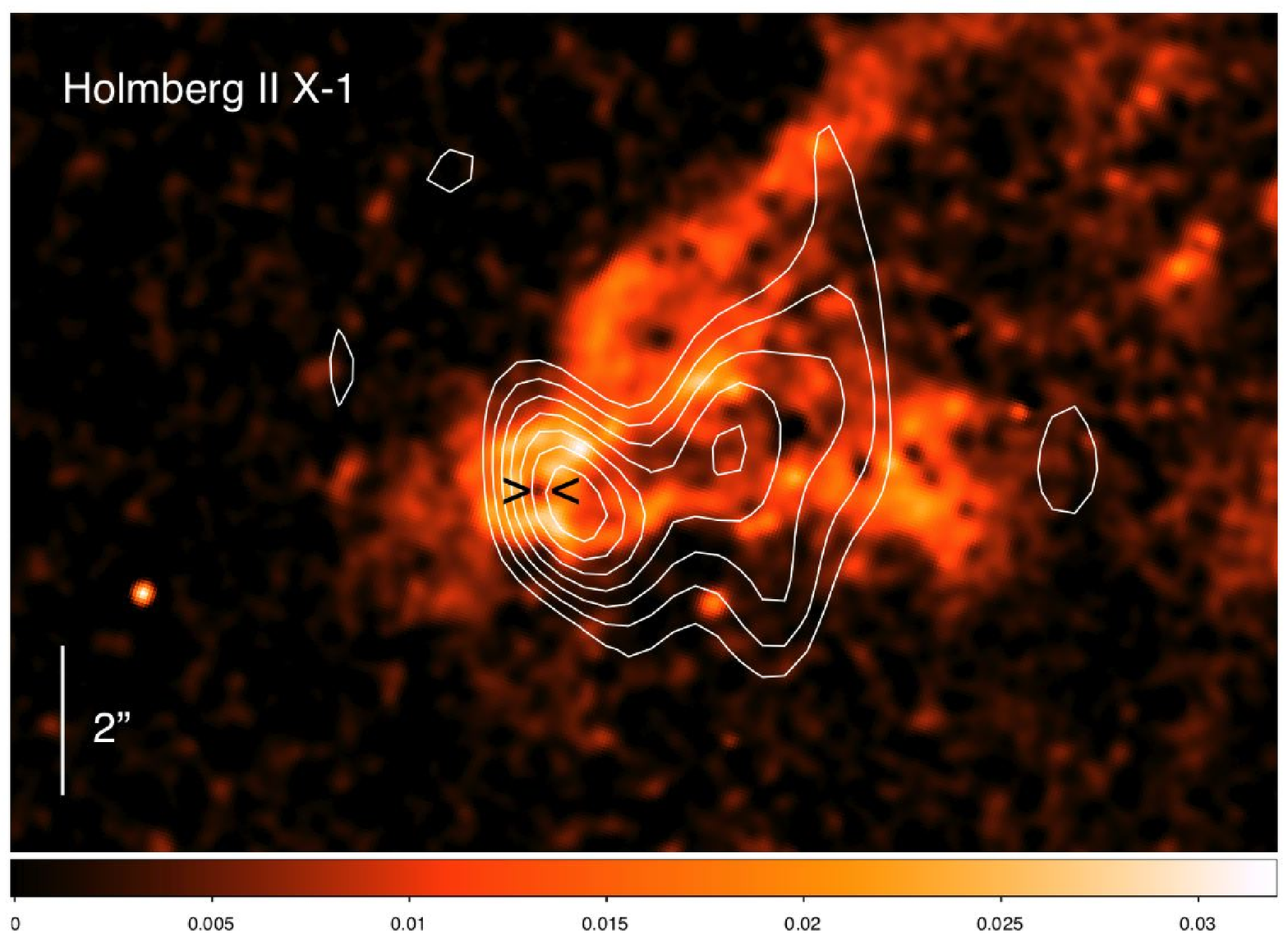}}
\caption{{\bf Left:} The 5 GHz VLA B-array image of {\bf  Ho II X-1} overlaid on the He {\sc ii}  HST image. Contours represent radio emission at levels of 3, 4, 5, 6 and 7 times the rms noise level of 14~$\mu$Jy beam$^{-1}$. The peak brightness is 114~$\mu$Jy beam$^{-1}$. The image was made using Robust=0 weighting and the resolution is $1\farcs5 \times 1\farcs0$ at PA=$34\degr$. {\bf Right:} The 8.5 GHz VLA C-array image of {\bf  Ho II X-1} overlaid on the H$\beta$  HST image. Contours represent radio emission at levels of 3, 4, 5, 6 and 7 times the rms noise level of 15 $\mu$Jy beam$^{-1}$. The peak brightness is 145 $\mu$Jy beam$^{-1}$. The image was made using Robust=-1 weighting and the resolution is $2\farcs36 \times 1\farcs75$ at PA=$3\degr$. The sign '$>\,<$' marks the X-ray position of the ULX. The North direction is up in both images.}
\label{HoII}
\end{figure*}

The radio nebula of Ho II X-1 was first detected at 1.4 and 5~GHz by \citet{miller}. Here, we have conducted observations of Ho II X-1 at 5 and 8~GHz in order to constrain the shape and spectrum of the radio nebula.  Figure~\ref{HoII}. shows the VLA images of Ho II X-1 overlaid on HST He {\sc ii} and H$\beta$ images from \citet{hoii}. On the left, the unweighted (Robust=0) 5~GHz B-array image is shown.  The right panel shows a Robust=$-$1 image made at 8~GHz using the C-array configuration. The asymmetric morphology of the nebula might indicate some outflows or ambient density gradient to the West. 

\begin{figure}
\begin{center}
\includegraphics[width=3.5in]{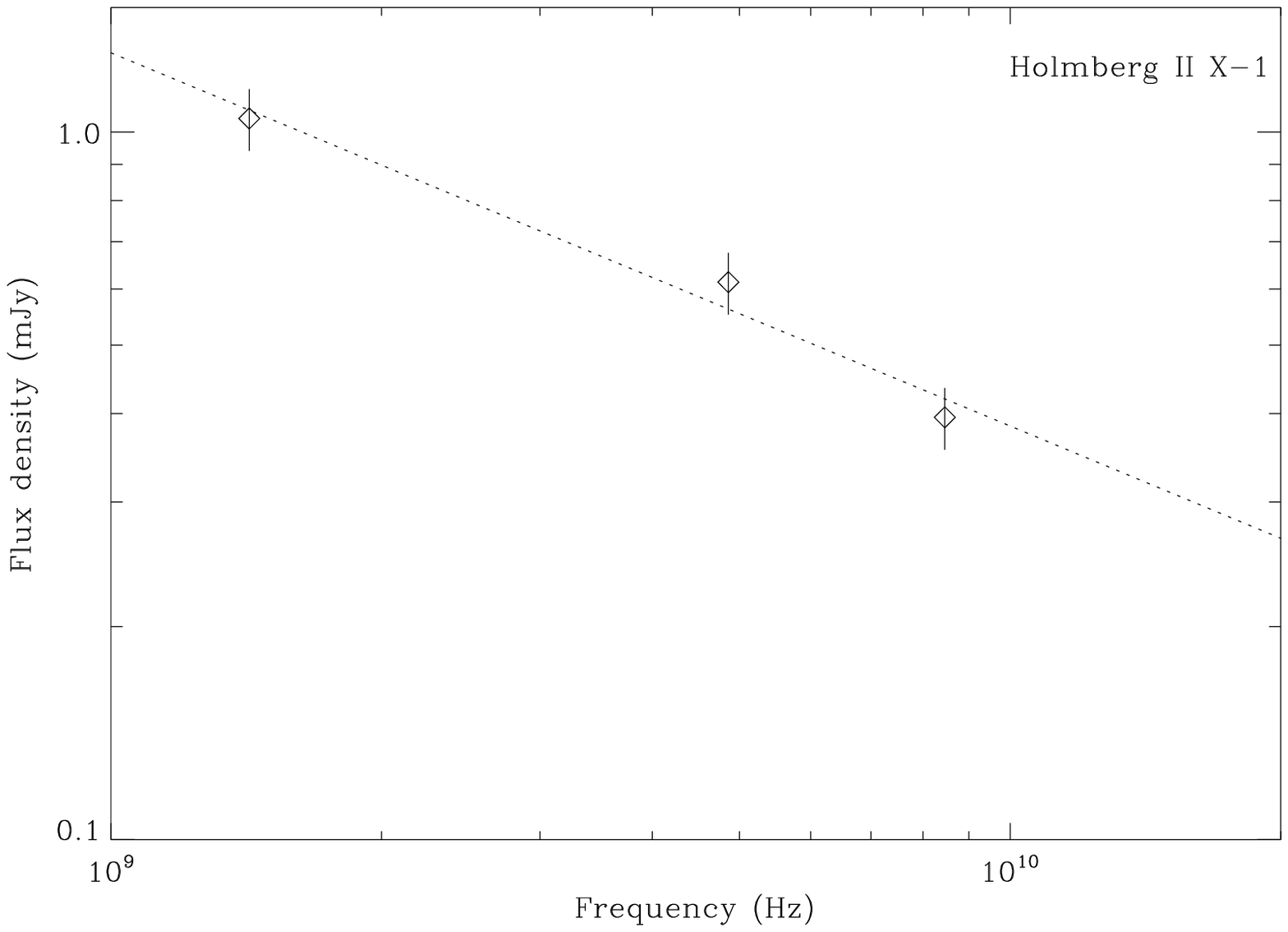}
\end{center}
\caption{The radio spectrum of the nebula of {\bf Ho II X-1}. The best fit spectral index is $\alpha=-0.53 \pm 0.07$ for a flux density, $S\propto \nu^{\alpha}$.}
\label{spix}
\end{figure}

Previously, the radio spectrum was not well constrained; \citet{miller} had high uncertainties on the flux at 5~GHz of Ho II X-1. However, we now have flux density measurements at three frequencies (1.4, 8 and 5~GHz, using the image from \citet{miller} at 1.4~GHz) and Figure~\ref{spix}. illustrates the fitted three-point spectral index of $\alpha=-0.53 \pm 0.07$, ($S\sim \nu^{\alpha}$). The radio spectrum is consistent with optically thin, synchrotron emission, further constraining the nebular nature of the radio counterpart of this ULX nebula. The integrated flux density of the nebula is 1.05$\pm$0.10~mJy, 613$\pm$61~$\mu$Jy and 395$\pm$40~$\mu$Jy at 1.4, 5 and 8~GHz.

\subsection{Swift/XRT and EVN observations of IC342 X-1}

\subsubsection{VLBI observation}

We conducted simultaneous X-ray and radio measurements of the VLA core (see Sec. 2.2.1.) to test its compactness in order to estimate the mass of the BH (see Sec. 3.4) with the minimum uncertainty. To check whether the VLA core is consistent with a compact jet at 10 mas scale, we carried out European VLBI Network (EVN) observations (EVN program code: EC032) at 1.6 GHz. The 12-hour observations were accommodated on June 15, 2011 from 03:30 to 15:30 (UT), simultaneously with the Swift/XRT observations. The participating VLBI stations were  Effelsberg (Germany), Jodrell Bank Lovell Telescope, Cambridge (United Kingdom), Medicina, Noto (Italy), Toru\'n (Poland), Onsala (Sweden),
Urumqi (P.R. China), Svetloe, Zelenchukskaya, Badary (Russia) and the phased array of the Westerbork Synthesis Radio Telescope (WSRT; The Netherlands). The aggregate
bitrate per station was 1024~Mbps. There were eight 8~MHz intermediate frequency channels (IFs) in both left and right circular polarisations.

The source was observed in phase-reference mode. This allowed us to increase the coherent integration time spent on the target source and thus to improve the sensitivity of the observations. Phase-referencing involves regularly interleaving observations between the target source and a nearby, bright, and compact reference source. The delay, delay rate, and phase solutions derived for the phase-reference calibrator (J0344+6827) was interpolated and applied to the target within the target-reference cycle time of 5~min. The target source was observed for 3.5-min intervals in each cycle.

AIPS was used for the VLBI data calibration following standard procedures \citep{dia}. The visibility amplitudes were calibrated using system temperatures and antenna gains measured at the antennas. Fringe-fitting was performed for the calibrator sources using 3-min solution intervals. The calibrated visibility data were exported to the Caltech Difmap program \citep{shep} used to make a naturally weighted VLBI image. The achieved 1-$\sigma$ rms noise level was 7 $\mu$Jy beam$^{-1}$ and we did not detect any source above the 3-$\sigma$ noise level in a field of view of 700 $\times$ 700 mas. Therefore, the VLA core is likely to be a clump of emission from the nebula. Our EVN observation places a 3-$\sigma$ upper limit on the flux density of $F_\nu \leq$ 21 $\mu$Jy and on the luminosity of the putative compact jet of  $\leq 1.9\times10^{33}$ erg s$^{-1}$.

\subsubsection{Swift observation}

The Swift X-ray Telescope (XRT) obtained 9394 seconds of good exposure in its photon-counting (PC) mode on June 15, 2011, from 02:57:31 to 17:47:47 (UT). We retrieved level two event files and used the default data screening methods as described in the XRT user's guide\footnote{http://heasarc.nasa.gov/docs/swift/analysis/}.  We extracted an X-ray spectrum for the source using a circular extraction region with a radius of 20 pixels (corresponding to 90\% of the PSF at 1.5 keV); background was estimated from a nearby circular region with a radius of 40 pixels and subtracted. We fitted the X-ray spectrum using the XSPEC \citep{arn} spectral fitting tool and the swxpc0to12s6\_20070901v011.rmf \footnote{http://heasarc.nasa.gov/docs/swift/proposals/swift\_responses.html} response matrix.

Fitting the spectrum in the 0.3-10 keV band, with an absorbed power-law, leads to a good fit with $\chi^2/{\rm DoF} = 14.7/25$ with a photon index of $\Gamma=1.57^{+0.29}_{-0.26}$ and an equivalent hydrogen absorption column density of $N_{\rm{H}}=4.5^{+2.0}_{-1.5} \times 10^{21}$ cm$^{-2}$. The absorbed flux in the 0.3-10 keV band was 1.97$\times 10^{-12}$ erg cm$^{-2}$ s$^{-1}$. The source clearly has a hard X-ray spectrum with a flux of $2.5 \times 10^{-12}$ erg cm$^{-2}$ s$^{-1}$, slightly lower than any of the previous XMM observations. The column density is consistent within the errors with the XMM values from \citet{feng}. The unabsorbed flux is 2.67 $\times 10^{-12}$ erg cm$^{-2}$ s$^{-1}$, corresponding to an unabsorbed luminosity of $4.86 \times 10^{39}$ erg s$^{-1}$ in the 0.3-10 keV band at a distance of 3.9 Mpc.

\subsection{ATCA observations of NGC5408 X-1}

The radio nebula of NGC5408 X-1 was the first detected radio counterpart of a ULX \citep{phil}. Later, it was confirmed that it is an extended source \citep{cornelia}. We obtained deep high-frequency ATCA CABB (Compact Array Broadband Backend) \citep{cabb} observations to better constrain the morphology and the high-frequency part of the radio spectra of the nebula at 5.5, 9, and 18~GHz. 

We observed NGC5408 X-1 with the CABB-upgraded ATCA in configuration 6D (baselines up to 6~km) on 2009 Aug 23 (program code: C1159). The data were obtained simultaneously at 5.5 \& 9~GHz and at 17 \& 19~GHz with 12~h on-source integration time for both sets (Table 1.). We observed in phase-reference mode; the phase calibrator was 1424-418 and the primary calibrator was PKS 1934-638. The data reduction was performed using the {\sc miriad} software package \citep{miriad}. We combined the 17~GHz and 19~GHz images in order to enhance sensitivity.

\begin{figure}
\begin{center}
\includegraphics[width=3.5in]{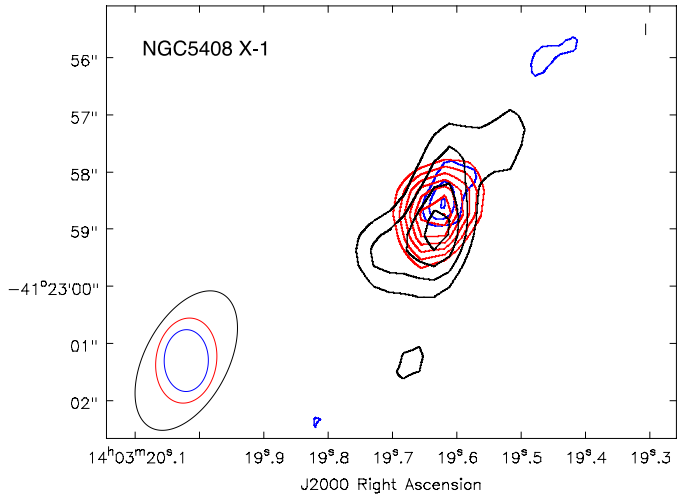}
\end{center}
\caption{The 5.5 (red), 9 (black) and 18 (blue) GHz ATCA 6D-array image of {\bf  NGC5408 X-1}. Contours represent radio emission at levels of 3, 4, 5, 6 and 7 times the rms noise level of 23, 17 and 13 $\mu$Jy beam$^{-1}$, respectively. The peak brightnesses are 208, 110, 66 $\mu$Jy beam$^{-1}$. The image was made using uniform weighting at 5.5~GHz and natural weighting at 9 and 18~GHz and the resolutions are $1\farcs5 \times 1\farcs0$ at PA=$-9\degr$, $2\farcs6 \times 1\farcs5$ at PA=$-28\degr$, $1\farcs0 \times 0\farcs8$ at PA=$-1\degr$ at 5.5, 9 and 18~GHz, respectively.}
\label{overlay}
\end{figure}

Fig.~\ref{overlay} shows the ATCA images of the nebula surrounding NGC5408 X-1 at 5.5, 9 and 18~GHz. The 5.5-GHz image was uniformly weighted, and the 9 and 18-GHz maps were naturally weighted in order to achieve the best sensitivity. Plotting the geometric mean beam size vs. the flux density, \citet{cornelia} found that the turnover indicates that the radio emission is resolved with an angular size of 1.5--2.0~arcsec. Our new radio images show that the size of the nebula is consistent with the previously estimated angular size.  At 9 and 18~GHz the source extent is only slightly larger than the corresponding beam sizes, but the elongated contours suggest that the nebula is resolved. This is typical for a weak, optically thin, steep-spectrum source, i.e.\ going towards higher frequencies, one gains resolution while the relative sensitivity decreases. 

\begin{figure}
\begin{center}
\includegraphics[width=3.5in]{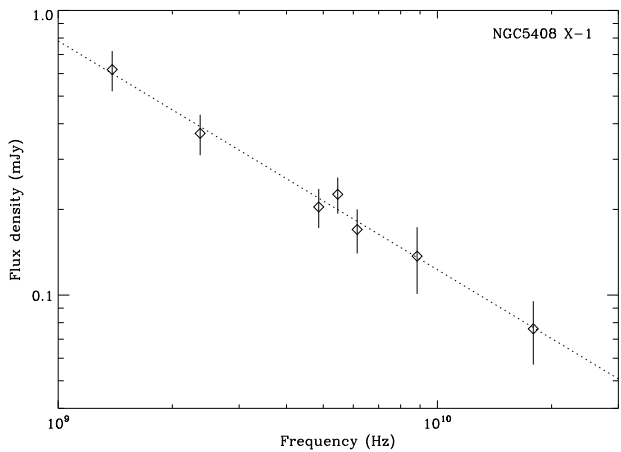}
\end{center}
\caption{The radio spectrum of the nebula of {\bf NGC5408 X-1}. The best fit spectral index is $\alpha=-0.8 \pm 0.1$ for a flux density, $S\propto \nu^{\alpha}$.}
 \label{spix2}
 \end{figure}

We used all available measurements to fit the radio spectrum. Fig.~\ref{spix2} shows previous measurements \citep{cornelia} and our new ones; combined they cover the 1.4 -- 18~GHz frequency range. Our new flux densities are 226$\pm$33, 137$\pm$36, and 76$\pm$20~$\mu$Jy at 5.5, 9, and 18~GHz, respectively. The fitted radio spectral index of NGC5408 X-1 is $\alpha=-0.8 \pm 0.1$.  This is the best constrained spectrum of a ULX nebula to date and is consistent with previous results suggesting optically thin synchrotron emission \citep{soria,cornelia}.

\begin{figure}
\begin{center}
\rotatebox{0}{
\includegraphics[width=3.5in]{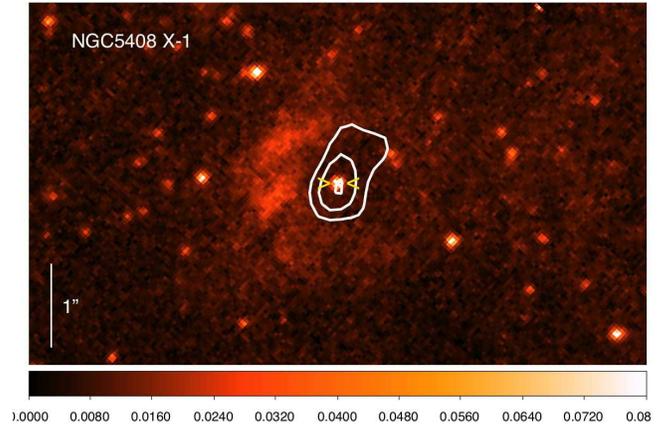}}
\end{center}
\caption{The 18 GHz ATCA 6D-array, naturally-weighted image of {\bf  NGC5408 X-1} overlaid on the  [O {\sc iii}]  HST image. The sign '$>\,<$' marks the X-ray position of the ULX. The North direction is up.}
\label{overlay2}
\end{figure}

Fig.~\ref{overlay2} shows the ATCA image at 18~GHz, overlaid on the HST image of the optical nebula of NGC5408 X-1 of \citet{fab}. The HST filter was centered on the forbidden [O {\sc iii}] nebular emission line (Table~2). The optical image shows a one-sided shell-like structure displaced from the ULX which might be due to geometrical effects like limb brightening towards the NE. We find that the radio emission at 18~GHz originates "inside" the optical nebula, however it is possibly an effect of the steep radio spectrum.

\subsection{ESO VLT observations of NGC5408 X-1}

As one can gain information about the ionization process from the line flux ratios of H$\alpha$ vs the forbidden sulphur lines, we conducted VLT observations.  VLT FORS-2 observations of NGC5408 X-1 were obtained on 12 April 2010 using the GRIS\_1200R grism with a slit width of $1.0\arcsec$ covering the spectral range 5750$-$7310 \AA\ with dispersion 0.38 \AA\ pixel$^{-1}$ and spectral resolution $\lambda/\Delta\lambda=2140$ at the central wavelengths, respectively.  The observation block (OB) consisted of three 849~s exposures with a 12 pixel offset along the spatial axis between successive exposures.  CCD pixels were binned for readout by 2 in both the spatial and spectral dimensions. The average seeing for our new observations was 0.62 arcsecond. Data reduction steps can be found in \citet{cseh}. Here we use a trace width of 8 pixels corresponding to $2\arcsec$ in order to accept a large fraction of the nebular emission.

\begin{figure}
\begin{center}
\includegraphics[width=3.5in]{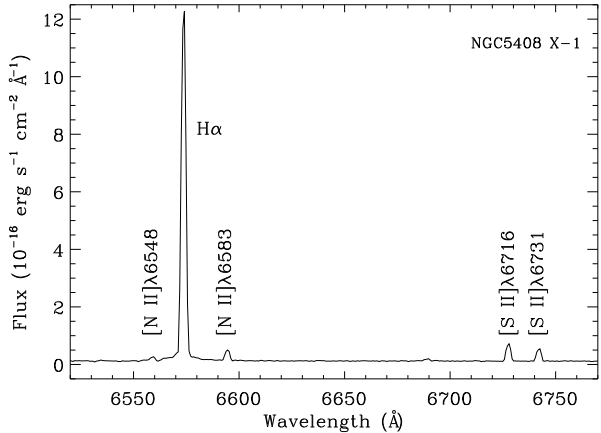}
\end{center}
\caption{VLT optical spectrum of the nebula around {\bf NGC5408 X-1}. The dereddened flux is plotted vs. wavelength and no redshift correction was applied.}
\label{vlt}
\end{figure}

Fig.~\ref{vlt} shows the optical spectrum of NGC5408 X-1. This portion of the optical spectrum shows the forbidden sulphur and nitrogen lines and the $H\alpha$ line. The lines are at wavelengths of 6558.9 ([N {\sc ii}]),  6573.7 (H$\alpha$), 6594.4 ([N {\sc ii}]), 6727.6  ([S {\sc ii}]), and 6742.0  ([S {\sc ii}]) in units of \AA. The corresponding fluxes are $0.46\pm0.02$, $33.1\pm0.1$, $1.07\pm0.02$, $1.57\pm0.01$, $1.15\pm0.01$ in the units of $10^{-16}$erg s$^{-1}$ cm$^{-2}$. The Gaussian FWHMs of the lines represent the instrumental resolution of $\sim$2.4 \AA.

\section{Discussion}

\subsection{The optical nebula around IC342 X-1}

In this section we use models developed for expanding bubbles, to calculate parameters of the nebula based on the observed optical properties. We follow the formalism developed by \citet{weaver} and \citet{ostriker} describing the hydrodynamic structure of a bubble. These formulae are valid for shock-dominated sources, i.e.\ jet or wind inflated bubbles, but not for photoionized bubbles. As IC342 X-1 is considered to be a shock-dominated source \citep{pakull2,rob,abol}, we can apply the well-known self-similar expansion law as a function of time, $t$:
\begin{equation}
R = \left(\frac{125}{154 \pi}\right)^{1/5} \left(\frac{L_{tot}}{\rho_0} \right)^{1/5} t^{3/5}
\end{equation}
where $L_{tot}$ is the mechanical luminosity (corresponding to a jet or a wind or an initial explosion energy), $R$ is the radius of the bubble expanding with velocity $v_{exp}=\dot{R}$ into the ISM. The mass density $\rho_0$ is assumed to be constant and $\rho_0 = \mu m_p n$, where $\mu=1.38$ is the mean atomic weight, $m_p$ is the proton mass and $n$ is the hydrogen number density. The characteristic age of the bubble is $\tau=3R/5v_{exp}$. The kinetic energy carried by the swept-up mass in the expanding shell is $E_k = \frac{15}{77}L_{tot} t$, while the energy emitted (the cooling) by the fully radiative shock expanding into the ISM is $E_{rad} = \frac{27}{77} L_{tot} t$. The thermal energy of the gas between the reverse shock and the swept-up shell is $E_{th} =\frac{5}{11} L_{tot} t$. Thus, the total energy is $E_{tot}=E_k+E_{rad}+E_{th}=L_{tot} t$. 

\citet{fengIC} found that the bright main body of the nebula has an angular diameter of about $6\arcsec$ in the $H\alpha$ image.  \citet{pakull2,fab00,fengIC} report an additional elongated structure to the South-West.  Considering the entire structure of the bubble, we estimate the volume of the optical nebula by taking a sphere with a diameter of $\sim10\arcsec$, which corresponds to $\sim$190~pc at a distance of 3.9~Mpc \citet{dist}.

The high flux ratio between the forbidden [S {\sc ii}] lines at 6,716 \AA \, and 6,732 \AA \, and the Balmer H$\alpha$ line indicates the presence of shock-ionized gas. \citet{abol} infer a shock velocity of $v_{exp} \simeq 20 - 100$ km s$^{-1}$, however, line ratios of a standard library of radiative shock models, e.g.\ He {\sc ii} vs. H$\beta$ is 0.036$\pm$0.015, suggest a shock velocity of $\simeq100$ km s$^{-1}$ \citep{allen}. Using this velocity, we find the characteristic age of the bubble is $ \tau=5.6 \times 10^5$~yr. \citet{fengIC} found that the color-magnitude diagram suggests that the minimum stellar age in the environment of the ULX is 10~Myr. The characteristic age of the nebula is much shorter and might suggest that the nebula formation is not related to the formation of the central BH or BH progenitor, instead it might represent the actively accreting phase of the binary.

Recalling the scaling of the total radiative flux and the flux in the Balmer lines \citep{dopita, abol}\footnote{We note, that there is a typo in Eq. 3.3 of \citet{dopita} (and consequently in Eq. 1 of \citet{abol}), pointed out to us by M.~W.~Pakull. The numerical factor should read $1.14 \times10^{-3}$ rather than $2.28 \times 10^{-3}$ as the maximum radiative flux is $\frac{1}{2} \rho v^3$ and not $\rho v^3$.}:
\begin{equation}
L_{H\beta}= 6.53 \times 10^{-3} \left(\frac{v_{exp}}{\rm{100\, km\,s^{-1}}} \right)^{-0.59}L_{rad}
\end{equation}
So, the total shock power represented as radiative losses is $L_{rad}=1.2 \times 10^{39}$~erg s$^{-1}$; using the $H\beta$ flux of \citet{abol} of $4.3\times10^{-15}$erg s$^{-1}$ cm$^{-2}$. As a consequence, the total mechanical luminosity is $L_{tot}=\frac{77}{27}L_{rad}=3.4\times 10^{39}$~erg s$^{-1}$, and the total kinetic power carried by the swept-up mass is $L_k=6.6\times 10^{38}$~erg s$^{-1}$. Similarly, the internal (thermal) luminosity is $L_{th}=1.5 \times 10^{39}$~erg s$^{-1}$. The energy, we see at t=$\tau$ is $E_{tot}=L_{tot}\tau=6.0 \times 10^{52}$~erg. From Eq.~1 one can derive that $n=1.0$~cm$^{-3}$ by substituting $R, \tau,L_{tot}$. The optical swept-up mass is then $M=\mu n m_p V=2.4 \times 10^{38}$~g or  $M=1.2 \times 10^{5}$ M$_{\odot}$.

IC342 X-1 is sometimes considered as a SN remnant \citep[eg.]{rob}. Here we intend to show that the total energy content does not depend significantly on the interpretation of the origin of the bubble. A SNR in the pressure-driven snowplow stage -- ie. radiative dominant phase following the adiabatic phase -- has an initial explosion energy, $E_i$\citep{cioffi}:
\begin{equation}
E_i=6.8 \times 10^{43}\left(\frac{R}{\rm{pc}}\right)^{3.16}\left(\frac{v_{exp}}{\rm{km\,s^{-1}}}\right)^{1.35}\left(\frac{n}{\rm{cm^{-3}}}\right)^{1.16}\rm{erg} 
\end{equation}
We have treated the metallicity correction factor, $\zeta_m^{0.161}$, as 1 for clarity \citep{cioffi,thornton}. Substituting $R,\, v$ -- and taking $n=1.0$ cm$^{-3}$ (see above), we obtain $E_i = 6.0 \times 10^{52}$~erg. We note that a similar formula of \citet{chev} provides $E_i = 5.0 \times 10^{52}$~erg. This initial energy is remarkably high, although somewhat expected as a single/simple SNR will not remain visible once it has expanded beyond $R\simeq 100$~pc with a canonical $E_i\simeq10^{51}$~erg \citep{mato,rob}. The estimated energy is in good agreement with the total energy content of the bubble, $E_{tot}$, within model uncertainties.  We note that an SNR nature for the nebula around IC 342 X-1 is strongly challenged by its high shock velocity coupled with large size \citep{pg08}.

\subsection{The radio nebula around IC342 X-1}

When radiation is via synchrotron emission, one can assume equipartition between energy of relativistic particles and the magnetic field. We calculate the minimum total energy of the radio nebula that corresponds to equipartition \citep{longair}:
\begin{equation}\label{emin}
E_{min}= 3 \times 10^{13} \eta^{4/7} \left(\frac{V}{\rm{m^3}}\right)^{3/7}  \left(\frac{\nu}{\rm{Hz}}\right)^{2/7}  \left(\frac{L_{\nu}}{\rm{W\,Hz^{-1}}}\right)^{4/7}  \rm{erg}
\end{equation}
where $\eta - 1$ is the ratio of energy in protons to relativistic electrons, $V$ is the volume, $\nu$ is the observing frequency and $L_{\nu}$ is the synchrotron luminosity. As is customary, we do not account for relativistic protons, therefore $\eta-1=0$ \citep[eg.][]{fender}. Substituting the corresponding values of $V\simeq1.79 \times 10^{56}$ m$^{3}$, $\nu=5 \times 10^9$ Hz, $L_{\nu}=3.64 \times 10^{18}$ W Hz$^{-1}$ and a filling factor of unity, we find the energy required to power the nebula is $E_{min}=9.2 \times 10^{50}$ erg. This suggests that the radio-emitting material carries a fraction $\sim10^{-2}$ of the initial energy. For comparison, the energy carried by mildly relativistic material in normal Type Ic supernovae has been suggested to be at most 10$^{-4}$ \citep{zsolt}, while for the jet inflated bubble around the extragalactic microquasar S26, this fraction is a few times 10$^{-3}$ \citep{s26}. These fractions suggest that most of the energy is stored in protons, nuclei, and non-relativistic bulk motion.

We calculate the magnetic field strength corresponding to the minimum energy condition \citep{longair}
\begin{equation}
B_{min}= 1.8 \times 10^{10} \left(\frac{\eta L_{\nu}}{V}\right)^{2/7} \nu^{1/7} \,\,\,\rm{\mu G}
\end{equation}
where the units and inputs are the same as above. We obtain $B_{min}=7.4\, \mu$G.

When energy loss is due to synchrotron radiation, the lifetime of an electron is \citep{longair,valeriu}
\begin{equation}
\tau_{sy}=2.693 \times 10^{13} \left(\frac{\nu}{\rm{Hz}}\right)^{-1/2}\left(\frac{B_{min}}{\rm{\mu G}}\right)^{-3/2}\,\,\, \rm{yr}
\end{equation}
Substituting $\nu=5 \times 10^9$ Hz and $B_{min}=7.4\, \mu$G, we find $\tau_{sy}=18.8$~Myr. So, consistently, the cooling time-scale is $\sim$34 times longer than the age of the bubble and $\sim$2 times higher than the minimum age of the ULX stellar environment.

We note that the value of $\eta$, the energy ratio of relativistic protons to electrons, can be estimated from the high-energy part of SNR spectra \citep{fermi,veritas}. Assuming IC342 X-1 has a ratio similar to that found for a scenario in which the high-energy gamma-rays from the Tycho SNR  are produced by leptons \citep{veritas}, then $\eta-1=10^2$ and the minimum energy obtained from Eq. \ref{emin} is increased by a factor of  $\eta^{4/7}=100^{4/7}=13.9$. This would mean that $E_{min}/E_{tot}=0.2$.  Instead, if the gamma-ray emission from Tycho is from hadrons, then $\eta-1=2.5\times10^{3}$ and $E_{min}/E_{tot}$ would be $ \simeq1.3$. Thus, if the high-energy gamma-ray emission of shock inflated bubbles originates from the same electron distribution that produces the radio, then via the $\eta$ parameter, the value of $E_{min}$ could increase, but interestingly, does not violate the total energy obtained from the optical.

\subsection{The radio and optical nebulae of Ho II X-1 and NGC5408~X-1}
\subsubsection{Ho II X-1}

Our new, higher resolution VLA radio observations clearly reveal that the morphology of the radio nebula follows the structure of the optical one (Fig.~\ref{HoII}). The radio nebula is resolved with a size of 5.5" $\times$ 2.7", corresponding to $\sim$ 81 pc $\times$ 40 pc. One could argue that this morphology reflects either a jet activity or an outflow. However, \citet{pakull3,hoii} showed, that the nebula of Ho II X-1 is consistent with photoionization by the central compact source. On the other hand, a complex velocity structure in the inner regions of the optical nebula indicates the impact of the central object also in the form of winds or jets \citep{lehmann}. As the [S {\sc ii}] vs. H$\alpha$ ratio is $\ll 1$ \citep{abol}, i.e.\ collisional excitation of the nebula is negligible, the morphology probably reflects an outflow rather than a well-collimated jet. This outflow is either relatively weaker than a jet -- thus preventing shock-ionization -- or the outflow is more isotropic than collimated. We note that weakly collimated outflows, in addition to jets, have been directly observed in SS433 with VLBI \citep{zsolt2}. In addition, the asymmetry of the outflow is probably due to the fact that the nebula is density bounded to the East and South of the central object \citep{pakull3,hoii}.

Furthermore, the minimum energy required to power the radio nebula of Ho II X-1 is $2.6 \times10^{49}$~erg with a magnetic filed strength of $13\mu$G and the synchrotron lifetime is $\tau_{sy}=25$ Myr \citep{miller}. This energy requirement is a factor of 35 less than needed for IC342 X-1, a shock-dominated source, which might also support a weak or uncollimated outflow.

Interestingly, \citet{fab0} studied an X-ray state transition of Ho II X-1 and found that it is difficult to interpret the thermal component of the X-ray spectra as disk emission or thermal Comptonization. On the other hand, this thermal component might be linked to the complex velocity structure of the nebula, ie. it might be due to a disk wind that results in a complex velocity structure of the nearby environment. We note that disk wind signatures have been found for GRS1915+105 \citep{grs} and SS433 \citep{fabrika}, i.e.\ for sources likely accreting above their Eddington-limit. Searching for relativistic disk lines and/or Fe K absorption line variation in the X-ray spectrum of ULXs might help disentangle whether some of these ULXs are similar to high-accretion rate Galactic binaries.

\subsubsection{NGC5408 X-1}

Fig.~\ref{overlay2}. shows a shell-like optical morphology of NGC5408 X-1 and a filled structure of the radio nebula with a maximum intensity near the ULX. The optical nebula is resolved, with a diameter of $\sim$ 2.5", corresponding to $\sim$ 60 pc \citep{fab}, which is much smaller than the optical nebula of IC342 X-1. Probably, the nebula is powered only by photoionization without any jet or outflow activity. Our optical spectra also suggest this, as the [S {\sc ii}] vs. H$\alpha$ line ratio is $\ll 1$ (Fig.~\ref{vlt}.). Furthermore, the energy needed to power the radio nebula of NGC5408 X-1 is $3.6 \times10^{49}$~erg with an equipartition field of $16 \mu$G and a synchrotron cooling time of $\tau_{sy}=20$ Myr \citep{cornelia}, which is, similar to Ho II X-1, a factor of 25 less than the energy needed to power the radio nebula of IC342 X-1.

\subsection{Upper limit on the mass of the BH in IC342 X-1}

Self-absorbed compact jets are ubiquitous in the X-ray hard states of GBHBs \citep{steph,robfender}. IC342 X-1 was found to have a hard X-ray spectrum in all available XMM-Newton and Chandra observations covering the period from 2001 to 2006 \citep{feng,fengIC} as well as in our recent Swift/XRT measurement (see Sec. 2.4.1.), thus the presence of compact jets is expected if the hard X-ray spectrum equivalent to the canonical hard state of GBHBs.

In addition, the morphology of the radio nebula is somewhat similar to the system SS433/W50 and may suggest a jet orientation along an axis slightly East of North \citep{fengIC}. As we pointed out earlier, an unresolved radio source is detected at the location of the ULX. This morphology might argue against the notion that the compact emission is the hot spot at the end of the jet pointing close to our line of sight. To test the compactness of this emission, we conducted VLBI measurements using the EVN (Sec 2.2) and we did not detect any source above the 3-$\sigma$ noise level, thus it is likely to be consistent with a clump of emission from the nebula and might be an effect of a steep radio spectrum.  We can set an upper limit on the flux of the putative compact jet of $1.9\times 10^{33}$ erg s$^{-1}$, using our 3-$\sigma$ noise level of 21 $\mu$Jy.

When black holes are in the hard state, i.e.\ their accretion is radiatively inefficient and perhaps advection dominated or jet dominated, a relationship holds between X-ray luminosity, radio luminosity and black hole mass \citep{merloni, falcke}.  This relationship, the fundamental plane of black holes, has been studied on a wide mass range from black hole X-ray binaries to low luminosity active galactic nuclei. Using the correlation with the minimum scatter \citep{elmar2}:
{\scriptsize
\begin{equation}\label{fp}
 \log \left(\frac{M}{\rm{M_{\odot}}}\right)=1.02^{-1}\left(1.59 \log \left(\frac{L_R}{\rm{erg\,\,\,s^{-1}}}\right)
 -  \log \left(\frac{L_X}{\rm{erg\,\,\, s^{-1}}}\right) - 10.15\right). 
\end{equation}
}  
and our radio and X-ray measurements, one can estimate the upper limit of the mass of the black hole in IC342 X-1. However, we must consider the intrinsic rms scatter in the measured fundamental plane relation of 0.12~dex within one $\sigma$. Substituting the upper limit of radio luminosity of the putative compact jet ($L=1.9 \times 10^{33}$ erg s$^{-1}$) and the simultaneously measured X-ray luminosity $L_{X}=4.86 \times 10^{39}$ erg s$^{-1}$, we estimate the mass of the black hole to be $M_{\rm{BH}}\leq(1.0\pm0.3) \times10^{3}$~M$_{\odot}$. This limit is valid only if IC 342 X-1 enters the canonical hard state.  We note that further observations could help place tighter constraints on the BH mass or help test if ULXs exhibiting hard X-ray spectra are, indeed, in the radiatively inefficient, hard X-ray state (see section 3.5.1).

\subsection{Comparison of IC342 X-1 with S26, IC10 X-1 and SS433}

In this section we compare IC342 X-1 to sources whose jet power and accretion rate has been estimated from extended emission in optical and/or radio. Table~\ref{fit} shows the main characteristics of the specific sources: the average X-ray luminosity ($L_X$), the jet power estimated from the environment ($Q_j$), the mass of the black hole ($M_{\rm BH}$), the total energy content ($E_{tot}$), the lifetime of the bubble ($\tau$), and the minimum energy estimated from the radio counterpart of the nebula ($E_{min}$). 

\begin{table*}
\begin{center}
\caption{Comparison with X-ray sources embedded in nebula\label{fit}}
\begin{tabular}{lcccccc}
\tableline\tableline
Name & $L_X$ (erg/s)  & $Q_j$ (erg/s) & $M_{\rm{BH}}$ (M$_{\odot}$) & $L_{tot}\tau=E_{tot}$ (erg)&$\tau$ (yr)&$E_{min}$ (erg)\\
\tableline
IC342 X-1 & $1.6\times10^{40}$ &$ 3.4\times10^{39} $&$ \lesssim1.0\times10^{3*} $&$ 6.0\times10^{52} $&$ 5.6\times10^{5} $&$9.2\times10^{50}$\\
\tableline
S26 & $6.2\times10^{36}$ & $5\times10^{40}$ & n/a &$3.16\times10^{53}$&$2\times10^{5}$&$10^{50}$\\
IC10 X-1 & $1.5\times10^{38}$&$1.27\times10^{39}$&23-34&$2\times10^{52}$&$5\times10^{5}$&n/a\\
SS433 & $\sim10^{36}$ & $2\times10^{38}$ &16  &$2\times10^{51}$&$2\times10^{5}$&$10^{49**}$\\
\end{tabular}
\tablecomments{The Table shows the average X-ray luminosity ($L_X$), the time-averaged jet power estimated from the environment ($Q_j$), the mass of the black hole ($M_{\rm BH}$), the total energy content($E_{tot}$), the lifetime of the bubble ($\tau$), and the minimum energy estimated from the radio counterpart of the nebulae ($E_{min}$). SS433 \citep{kirsh,begelman2,dubner,blundell,perez,fabrika}. S26 \citep{pakull,s26}.IC10 X-1 \citep{lozi,bauer}.  *Estimated using the fundamental plane.**This value was calculated using Eq.~4 using the parameters found by \citet{dubner}: a radius of $\sim$30~pc, a distance of 3~kpc, a radio flux density of 71~Jy at 1.4~GHz, and taking a spectral index of -0.48 between 85 MHz and 5 GHz.  We note that this value is two orders of magnitude larger than that quoted by \citet{dubner}, however consistent with the value of \citet{begelman2}.}
\end{center}
\end{table*}

The nebula around IC10 X-1 is sometimes considered to be the result of a hypernova event \citep{lozi}.  However, its size is much larger than allowed by the surface brightness versus size ($\Sigma$-D) relation for explosive events, i.e.\ supernovae \citep{yang}, suggesting that it is powered, instead, by the central BH \citep{bauer}. S26 is another outlier source in the $\Sigma$-D relation and it has been revealed that the bubble is powered by a jet \citep{pakull,s26}.  Therefore, we will consider IC10 X-1 as a black hole powered nebula. We note that ``black hole powered nebula'' means there is a significant and continuing outflow from the central black hole powering the nebula and does not argue against the formation of the black hole in a SN explosion.

Considering S26, the X-ray photon index is $\Gamma=1.4$ \citep{s26}, consistent with a source being in the hard state. However, no radio core was detected with a 3-$\sigma$ upper limit of 0.03 mJy, which is one third of the peak intensity of IC342 X-1; and the total jet power of S26 is an order higher than for IC342 X-1 (Table~3). Using the fundamental plane (Eq.~\ref{fp}), we obtain an upper limit on the mass of $M\simeq(1.0\pm0.4)\times10^{7}$ M$_{\odot}$. Given that $Q_j \propto \dot{M}$ in the hard state \citep{elmar0,elmar3} and assuming similar accretion efficiencies for IC342 X-1 and S26; then one can speculate on the basis of the ratios of the jet powers that the average mass accretion rate of S26 is $\sim$15 times higher than that of IC342 X-1. 

IC10 X-1 has an X-ray photon index of $\Gamma=1.83$, potentially being in the hard state \citep{bauer}. Using the fundamental plane and adopting the average X-ray luminosity of $1.5\times 10^{38}$ erg s$^{-1}$ and a mass of $\sim30$ M$_{\odot}$ (Table~\ref{fit}), the expected core radio flux is then $\sim10$ $\mu$Jy at a distance of 0.7 Mpc. Thus, the source could be detectable with e-MERLIN or the EVLA.

\subsubsection{Jet characteristics, accretion rates and efficiencies}

In this and the following sections we investigate the possibility that the nebula around IC342 X-1 is powered by a jet and the consequences regarding the jet properties. 

The elongated morphology of the nebula  and its shock-ionized nature are indicative of jet inflation. If the nebula is inflated by a jet, then the total power available in the bubble has to be equal with the time-averaged total jet power, ie. $Q_j$=$L_{tot}$, derived from the optical bubble \citep{pakull, s26}. We note that calculating the jet power using the minimum energy ($E_{min}$) derived from the radio bubble would lead to an underestimation. This could be either due to a mild deviation from equipartition, resulting in $E_{radio,tot}/E_{min}=10-100$ \citep[eg.][]{zsolt} or due to the fact that the kinetic power associated to the bulk motion is transferred to thermal ions also \citep{s26}.

In general, the total jet power is a constant fraction ($f$) of the available accretion power (however see \citet{mickael} for possible variation of $f$).  Thus, we can write
\begin{equation}\label{def}
Q_j=fQ_{acc}=f\dot{M}_{acc}c^{2},\,\,\, f<1 
\end{equation}
where $f$ is typically in the range of $10^{-3}$ to $10^{-1}$ \citep{falcke95,falcke99}.

Taking a constant rate of energy input that is characterized by the power of the jet, the rest-mass transport along the jet, $\dot{M}_{j}$, is \citep{kaiser}:
\begin{equation}\label{rmt}
\dot{M}_{j}=\frac{Q_j}{(\gamma-1)c^2}
\end{equation}
Taking a minimum Lorentz factor of $\gamma=2$ \citep{mirabel,rf}, we find $\dot{M}_{j}\simeq 6.0 \times 10^{-8}$ M$_{\odot}$ yr$^{-1}$ ($3.8\times10^{18}$ g s$^{-1}$). 

\citet{elmar3} found that the total jet power could be estimated from the flux of the compact jet as:
\begin{equation}\label{kin}
Q_j\simeq7.2 \times 10^{36}\left(\frac{L_{jet,radio}}{\rm{10^{30}\, erg\, s^{-1}}}\right)^{12/17} \,\,\, \rm{erg\, s^{-1}} 
\end{equation}
This relationship was found for FR I and FR II radio galaxies and scaled to Cyg X-1. (See \citet{gallo} and \citet{dave2} for the jet power estimation of Cyg X-1). Substituting the total jet power obtained from the optical nebula ($Q_j=3.4\times10^{39}$ erg s$^{-1}$), we find that the expected average luminosity of the putative radio core would be $6.1 \times10^{33}$ erg s$^{-1}$. Not detecting a compact jet with an upper limit of $1.9\times 10^{33}$ erg s$^{-1}$ could mean that either its flux is variable, and currently below our detection limit but with an average above, or the hard X-ray spectrum does not represent the canonical hard state of GBHBs.
 
In addition, it is possible to estimate the accretion rate, for hard state objects, based on the luminosity of the compact radio jet (\citet{elmar0,elmar3}; and references therein):
\begin{equation}\label{fr2}
\dot{M}\simeq4\times10^{17}\left(\frac{L_{jet, radio}}{\rm{10^{30}\, erg\, s^{-1}}}\right)^{12/17} \,\,\, \rm{g\,s^{-1}}
\end{equation}
For IC342 X-1, we obtain an upper limit, due to the fact that we do not detect a compact jet, of $\dot{M}\leq8.2\times10^{19}$ g s$^{-1}$ ($1.3\times10^{-6}$ M$_{\odot}$ yr$^{-1}$). This rate is close to the Eddington rate for a $100 M_{\odot}$ compact object. Comparing the accretion rate to the jet mass loss rate of $\dot{M}_{j}\simeq 3.8\times10^{18}$ g s$^{-1}$ leads to the conclusion that the jet efficiency is $f\geq0.046$. This would be a typical value for a jet efficiency and could be consistent with the assumption of \citet{elmar3} of f=10$^{-1}$ for obtaining the relationship of Eq. \ref{fr2}. 

Let us compare IC342 X-1 to the peculiar sources SS433 and GRS1915+105. The jet mass flow rate in SS433 is $\dot{M}_{j}=5\times10^{-7}$ M$_{\odot}$ yr$^{-1}$ and the available accretion rate is $\dot{M}=10^{-4}$ M$_{\odot}$ yr$^{-1}$ \citep{perez,fabrika}, so the jet efficiency is $f=5\times10^{-3}$, which is also within the typical range for black hole jets. Comparing the accretion rate of IC342 X-1 to SS433, we find $\dot{M}_{\rm{IC342 X-1}}$/$\dot{M}_{\rm{SS433}} \geq 10^{-2}$. In contrast, the accretion rate of IC342 X-1 might be an order of magnitude higher than the rate of GRS 1915+105 of $\sim10^{19}$ g s$^{-1}$ \citep[eg.][]{rushton}. It is interesting to note that if one takes the lifetime of the surrounding bubble, and assumes a constant accretion rate \citep{beg}, the compact object in SS433 could accrete $\sim$20~M$_{\odot}$. During the lifetime of the nebula, the compact object in IC342 X-1 could accrete only $\lesssim$1 M$_{\odot}$ . This comparison of accretion rates suggests that there is no need to invoke super-Eddington accretion for IC342 X-1 if the mass of black hole is above $\simeq$100 M$_{\odot}$.

\section{Conclusions}
We studied the radio and optical nebulae of three ULXs. One of these ULXs, IC342 X-1, is a newly discovered radio association. 

\begin{itemize}

\item We confirmed that the radio spectra of the nebulae surrounding Ho II X-1 and NGC5408 X-1 are consistent with optically thin synchrotron emission.

\item We estimated the energy needed to supply both the optical and radio nebulae around IC342 X-1. The energy needed, $6\times10^{52}$ erg, is $\sim2$ orders of magnitude higher than the explosion energy of a SN. By comparing the age of the bubble to the stellar environment, we found that the nebula is much younger, indicating that the nebula formation is not necessarily related to the formation of the black hole progenitor.

\item We estimated that the minimum energy needed to power the radio nebula of IC342 X-1 is $9.2\times10^{50}$ erg, at least an order of magnitude higher than that of Ho II X-1 and NGC5408 X-1. The fraction of energy carried by relativistic material is relatively high.

\item In addition to discovery of a radio nebula around IC342 X-1, we found a radio component unresolved on VLA scales, that was not detected with VLBI. This puts an upper limit on the flux density of a compact jet. The ULX was found with a hard X-ray spectrum in observations covering $\sim10$ years. Use of the 'fundamental plane' of accreting black holes, which is valid for hard state objects, would place an upper limit on the mass of the black hole in IC342 X-1 of $(1.0\pm0.3) \times10^{3}$~M$_{\odot}$.

\item According to the above properties of IC342 X-1, we argued that the nebula is possibly inflated by a jet. We found that the calculated time averaged jet power, jet efficiency, and available accretion power could be consistent with that the nebula is inflated by a jet.

\end{itemize}

In summary, the energetics of the surrounding nebula along with the possible jet properties and accretion rate support the idea that the nebula surrounding IC342 X-1 could be an inflated bubble driven by the jet from the central black hole.

\acknowledgments

We thank the referee, M. W. Pakull, for the constructive comments that improved the quality of the paper.  The research leading to these results has received funding from the European Community's Seventh Framework Programme (FP7/2007-2013) under grant agreement number ITN 215212 "Black Hole Universe". PK and FG acknowledge partial support from NASA grant NNX10AF86G.The National Radio Astronomy Observatory is a facility of the National Science Foundation operated under cooperative agreement by Associated Universities, Inc. Based on observations made with ESO Telescopes at the La Silla or Paranal Observatories under programme ID 385.D-0782(A). Based on observations made with the NASA/ESA Hubble Space Telescope, obtained from the data archive at the Space Telescope Institute. STScI is operated by the association of Universities for Research in Astronomy, Inc. under the NASA contract  NAS 5-26555. The Australia Telescope is funded by the Commonwealth of Australia for operation as a National Facility managed by CSIRO. The European VLBI Network is a joint facility of European, Chinese, South African and other radio astronomy institutes funded by their national research councils.

{\it Facilities:} \facility{VLA, VLT:Antu, ATCA, EVN, Swift, HST}.

\end{document}